\documentclass[paper]{acoust_arxiv}
\usepackage{bm,multirow,url}

\def\header{Real-time short-response selection}

\title{Real-time short-response selection for spoken dialogue systems}

\author{Hien Ohnaka$^{1}$\thanks{h.ohnaka@dis.sys.es.osaka-u.ac.jp},
Kazuyo Onishi$^{2,1}$, and Koichiro Yoshino$^{1}$}

\inst{$^{1}$ Graduate School of Engineering Science, The University of Osaka, \\
1-3 Machikaneyama-cho, Toyonaka, Osaka, 560-8531, Japan \\
$^{2}$ Nara Institute of Science and Technology, 
8916-5 Takayama-cho, Ikoma, Nara, 630-0101, Japan \\}

\abst{
To realize spoken dialogue systems that can interact naturally with users, 
it is important for the system to respond to user utterances at appropriate timings. 
To achieve such natural turn-taking, 
the system needs not only to predict appropriate response timings but also to mask delays caused by language generation and speech synthesis so that they are not perceived by the user.
We propose a real-time short-response selection task that can be integrated with existing turn-taking models to mask generation delay.
To address this task, the proposed model learns the mapping between dialogue context and short responses using contrastive learning.
When a turn-taking event is detected, the system selects a response from a set of pre-generated candidates.
The selected response is played while the main response is being generated.
Experimental results show that the proposed method learns to select short responses based on dialogue context, outperforming random and LLM-based baselines.
Moreover, integrating response selection maintained turn-taking prediction performance and added minimal processing overhead.
}

\kword{spoken dialogue systems, turn-taking, contrastive learning, voice activity projection}

\mladdress{The University of Osaka, 1-3 Machikaneyama-cho, Toyonaka, Osaka, 560-8531, Japan}
\class{3}

\begin{document}
\maketitle

\section{Introduction}
The quality of spoken dialogue systems has greatly improved with the development of deep learning.
In addition to context-aware response generation using large language models (LLMs)~\cite{llama}, 
text-to-speech (TTS) systems have also improved the naturalness and expressiveness of synthetic speech~\cite{tacotron,f5-tts}.
However, natural spoken dialogue requires not only high-quality responses and speech but also appropriate turn-taking between dialogue participants~\cite{turn_taking_survey}.
In modeling natural turn-taking, the system needs to determine the appropriate timing of speaker changes and use backchannels, which are brief responses that indicate understanding or active listening during the other speaker's utterance. 
The interval before these responses begin affects the impression of the responder~\cite{timing_social_connection,timing_subjective_eval}, and appropriate backchannels also help the other speaker continue speaking~\cite{backchannel_functions}.
Therefore, smooth interaction in spoken dialogues requires not only better response generation but also better temporal coordination.

In turn-taking modeling, several methods have been proposed to predict turn-taking events from speech signals or their transcriptions~\cite{dnn-based-eot-pred,dnn-based-event-pred,rtnet}.
More recently, lightweight and fast models with streaming processing have enabled real-time detection of these events~\cite{vap,multilingual_vap,backchannel_vap}.
In practical cascaded spoken dialogue systems, an LLM and a TTS system become the main bottlenecks before a response can begin.
Even if a turn-taking model quickly detects an appropriate response timing, the generation systems cannot start responding until the response speech has been generated.
One possible approach is to start response generation early enough to meet the predicted response timing~\cite{remdis,dialogue_prefetching}.
However, the earlier generation starts, the more this approach has to rely on partial automatic speech recognition (ASR) transcripts.
As a result, the response may not fully reflect the user's intended meaning~\cite{vap_evaluation}.
Moreover, when attempting to generate a response that sufficiently reflects the user's intent, 
there often remains a gap between the appropriate response timing and the time when the generation modules can actually output the response.

One approach to reducing this gap is to speed up the generation modules themselves~\cite{quantization-aware-training,bitnet,clarinet}.
However, speeding up generation through model compression or similar techniques may involve a trade-off with generation quality.
Another approach is to output a short utterance before the main response through a separate processing path.
Previous studies~\cite{Filledpause_response1,Filledpause_response2} have shown that outputting a pre-prepared filler such as ``Hmm...'' can improve the user's impression of the waiting time.
However, repeatedly outputting the same utterance may make the dialogue monotonous and may result in responses that do not fit the context.

In recent years, several approaches have been proposed to address this problem.
One is a method called preface-generation~\cite{context_aware_preface_generation}.
It predicts whether a response can be generated from partial recognition results of the user's utterance.
Once a response is considered possible, it generates a contextually appropriate response based on the available context, separately from the main generation pathway, and outputs it at the timing determined by the turn-taking model.
Another is an LLM-based filler selection strategy proposed by Liu et al.~\cite{llm_based_selection}.
This approach selects fillers based on partial recognition results of the user's utterance.
These approaches may select or generate more natural fillers than the naive filler output described above.
However, their performance remains dependent on the speed of the ASR model, the LLM, and the TTS system.

\begin{figure}[t]
  \centering
  \includegraphics[width=\linewidth]{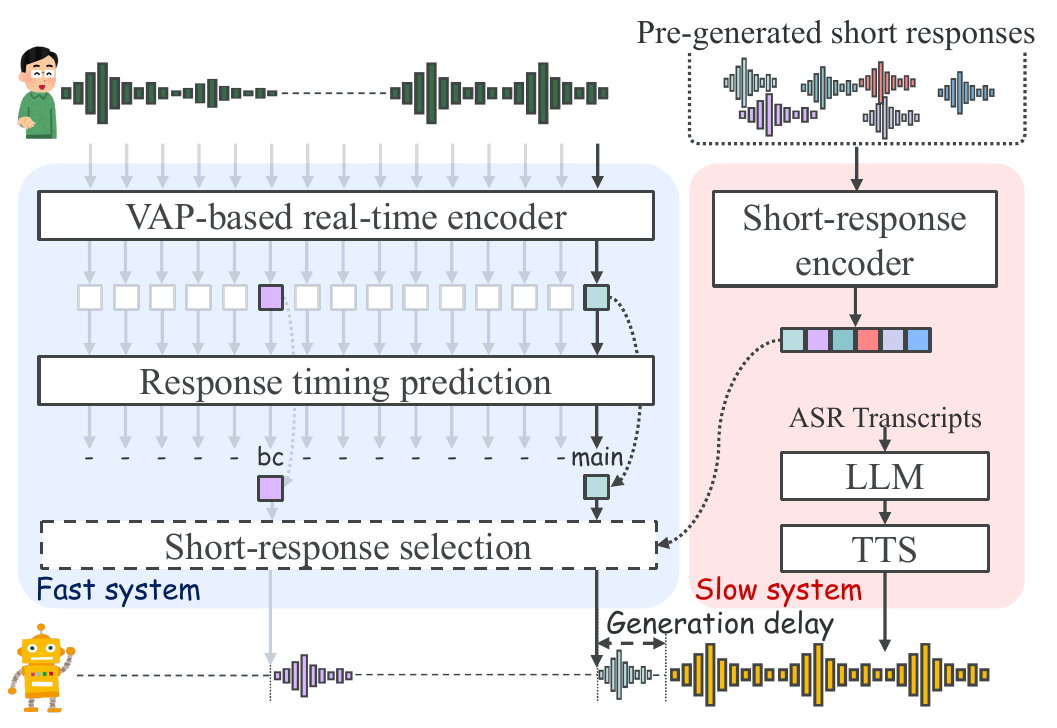}
  \caption{
    Overview of the proposed real-time short-response selection.
  }
  \label{fig:concept_proposed}
\end{figure}

In this study, we propose a real-time short-response selection task that predicts the appropriate timing for the system to start speaking while rapidly selecting contextually appropriate fillers or utterances intended to take the floor.
An overview of the proposed method is shown in Fig.~\ref{fig:concept_proposed}.
Our method first extracts hidden representations from a set of short responses.
Here, as shown in the upper right of Fig.~\ref{fig:concept_proposed}, the short responses are defined as brief utterances such as “I see” and “Um…” that can be produced before the main response. 
During inference, short responses are selected whenever an event is triggered by the continuously running turn-taking model.
The selected response is expected to match the latent dialogue context encoded in the turn-taking features.
Crucially, the selection process is independent of other modules, such as ASR and LLMs, and can therefore be performed with low latency.
The selected response can be output to mask the latency of the generation module and can also be used for backchannel events.
Our method is realized by adding a short-response encoder to the turn-taking model.
Specifically, in addition to the existing turn-taking objective, we introduce a contrastive objective between the dialogue context and short responses.
Our method can also be viewed as an extension of context–feedback embedding-based response ranking~\cite{contrastive_bc_selection} to streaming, timing-aware short-response selection.

In the evaluation experiments, we conducted a broad objective evaluation as an initial validation.
The results show that the method can select responses based on dialogue context, 
learn short-response selection without degrading timing prediction performance, 
and perform response selection with almost no additional overhead.
Finally, we discuss the current limitations of our method and directions for future improvement.

\section{Preliminary Definition}
\subsection{Spoken Dialogue Systems}
In this research, we assume a two-speaker dialogue setting, following previous work~\cite{multilingual_vap}.
We denote the user speech as $\bm{x}_u$ and the system speech as $\bm{x}_s$.
The two speech signals are treated as synchronized streams along the time axis.
For speaker $a \in \{u, s\}$, the speech history up to time frame $t$ is represented as follows:
\begin{align}
  \bm{x}_a^{\leq t} = \{x_a(t') \mid 0 \leq t' \leq t\}. \label{eq:streaming_input}
\end{align}

In a cascaded spoken dialogue system, 
an ASR module~\cite{whisper,asr_survey} first converts the user speech into text in a streaming manner as follows:
\begin{align}
  \hat{\bm{y}}_{\mathrm{asr}}(t)
  = f_{\mathrm{asr}}\!\left(\bm{x}_u^{\leq t}\right). \label{eq:streaming_asr}
\end{align}
Here, $f_{\mathrm{asr}}$ denotes a streaming ASR model~\cite{streaming_asr,streaming_nemotron} that outputs a partial text sequence $\hat{\bm{y}}_{\mathrm{asr}}(t)$.
In parallel with ASR, the turn-taking model sequentially predicts turn-taking events, 
such as response timing and backchannels, from the speech streams.
The event prediction is represented as follows:
\begin{align}
  \hat{\bm{e}}(t)
  = f_{\mathrm{tt}}\!\left(\bm{x}_u^{\leq t}, \bm{x}_s^{\leq t}\right). \label{eq:streaming_turntaking}
\end{align}
Here, $f_{\mathrm{tt}}$ denotes the turn-taking model, and $\hat{\bm{e}}(t)$ denotes the event prediction at time $t$.
Some turn-taking models depend on ASR to use textual information~\cite{rtnet,context_aware_preface_generation}, while others operate without ASR~\cite{vap,multilingual_vap}.
In this study, we focus on the latter type and treat it as a module that can operate in parallel with ASR.

Let $t_e$ denote the time frame when a turn-taking event is detected.
The response generation module then generates a response text using the recognition results obtained up to that time.
Let $p$ denote the prompt given to the LLM.
The generated response text is represented as follows:
\begin{align}
  \hat{\bm{y}}_{\mathrm{llm}}
  = f_{\mathrm{llm}}\!\left(\hat{\bm{y}}_{\mathrm{asr}}(t_e), p\right). \label{eq:large_language_model}
\end{align}
Here, $f_{\mathrm{llm}}$ denotes a large language model (LLM)~\cite{llama}.
Finally, a text-to-speech (TTS) system~\cite{tacotron,f5-tts} converts the generated response text into speech:
\begin{align}
  \hat{\bm{x}}_s
  = f_{\mathrm{tts}}\!\left(\hat{\bm{y}}_{\mathrm{llm}}\right), \label{eq:text_to_speech}
\end{align}
Here, $f_{\mathrm{tts}}$ denotes the TTS system, and $\hat{\bm{x}}_s$ denotes the response speech output by the system.

Cascaded spoken dialogue systems offer high flexibility because their modules operate independently.
As a result, each module in Eqs.~(\ref{eq:streaming_asr}), (\ref{eq:streaming_turntaking}), (\ref{eq:large_language_model}), and (\ref{eq:text_to_speech}) can be optimized separately and readily replaced at inference time.
However, the connections between these modules can introduce delay.
Another approach is an end-to-end spoken dialogue model~\cite{moshi}.
It avoids this problem caused by combining multiple modules by handling the process from speech input to speech output within a single model.
These systems have no explicit boundaries between modules.
Therefore, they may handle mismatches in processing timing across modules in an integrated manner.
For this reason, they have attracted attention from the perspective of natural turn-taking.
However, they require large-scale training data and computational resources~\cite{j-moshi}.
In addition, compared with cascaded systems, 
where individual modules can be independently replaced and optimized, controlling and customizing each component is less straightforward.

\subsection{Turn-taking modeling}
In this section, we focus on two-channel Voice Activity Projection (VAP)~\cite{vap,multilingual_vap}, which serves as the turn-taking model in our method.
Rather than directly predicting specific turn-taking events, VAP sequentially predicts the future speech activity of two speakers.

We consider the speech activity of speaker $a \in \{u,s\}$ over the next two seconds at time frame $t$.
In VAP, this interval is divided into four windows with durations $\bm{d} = (0.2, 0.4, 0.6, 0.8)$ seconds.
The binary speech activity label of speaker $a$ in the $k$-th window is defined as follows:
\begin{align}
  b_{a,k}(t) \in \{0,1\}. \quad k \in \{1,2,3,4\}
\end{align}
By concatenating the four labels for each of the two speakers, the future speech activity at $t$ is represented as follows:
\begin{align}
  \bm{b}(t)
  = (b_{u,1}(t),\ldots,b_{u,4}(t),b_{s,1}(t),\ldots,b_{s,4}(t))
  \in \{0,1\}^{8}
\end{align}
We regard these 8-bit combinations as a codebook $\mathcal{C}=\{0,1\}^{8}$.
A discrete class can then be assigned to the speech activity at each time step as follows:
\begin{align}
  c(t)=\operatorname{code}(\bm{b}(t)) \in \{1,\ldots,256\},
\end{align}
Here, $\operatorname{code}(\cdot)$ denotes a function that maps an 8-bit pattern to a class index.

\begin{figure}[t]
\begin{center}
\includegraphics[width=\linewidth]{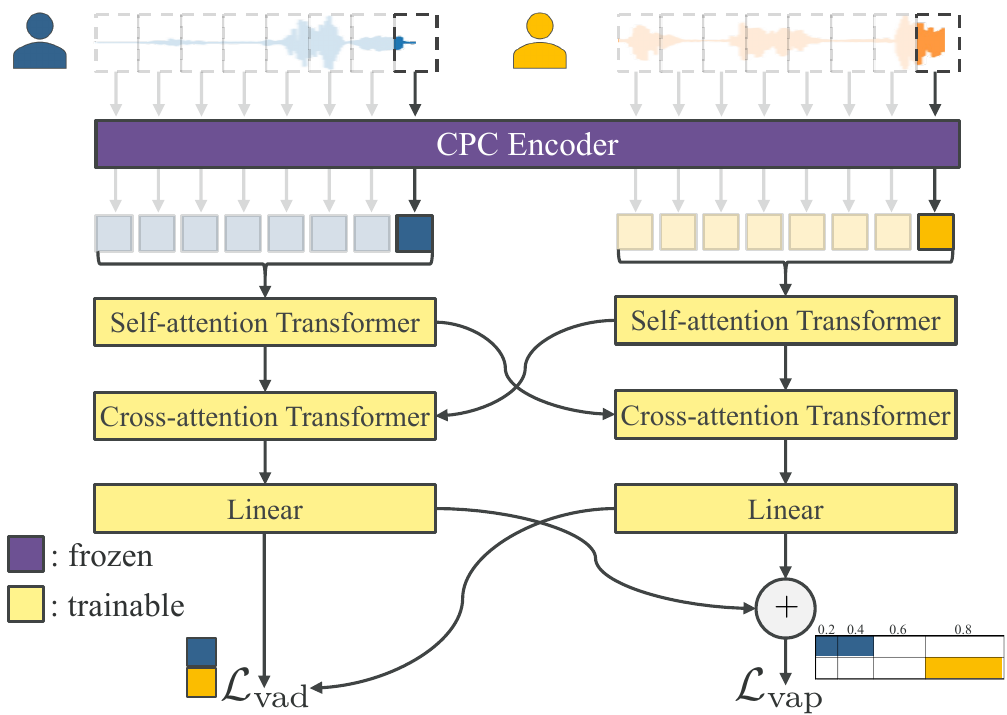}
\caption{Original VAP architecture used to initialize the turn-taking backbone. 
The contrastive predictive coding (CPC) encoder~\cite{cpc_encoder} is used as the feature extractor that converts each speaker's waveform into frame-level features.}
\label{fig:vap_architecture}
\end{center}
\end{figure}

VAP takes only the speech histories of the two speakers observed up to time frame $t$ as input.
Let $f_{\mathrm{vap}}$ denote the streaming model that extracts the two speakers' features.
Figure~\ref{fig:vap_architecture} shows the architecture of the model.
The streaming model consists of self-attention and cross-attention Transformer blocks~\cite{transformer}.
\begin{align}
  \bigl(\bm{z}_u(t),\bm{z}_s(t)\bigr)
  &= f_{\mathrm{vap}}\!\left(\bm{x}_u^{\leq t},\bm{x}_s^{\leq t}\right), \\
  \bm{z}(t)
  &= \bm{z}_u(t)+\bm{z}_s(t).
  \label{eq:vap_features}
\end{align}
Here, $\bm{z}_u(t)$ and $\bm{z}_s(t)$ denote the user and system features, respectively.
Let $h_{\mathrm{vap}}$ and $h_{\mathrm{vad}}$ denote the VAP state head and the voice activity detection (VAD) head, respectively.
The model solves this multi-class classification problem in a streaming manner as follows:
\begin{align}
  \hat{\bm{p}}_{\mathrm{vap}}(t)
  &= h_{\mathrm{vap}}\!\left(\bm{z}(t)\right)
  \in [0,1]^{256}, \label{eq:vap_probs} \\
  \quad \sum_{c=1}^{256}\hat{p}_{\mathrm{vap},c}(t)&=1.
\end{align}
Here, $h_{\mathrm{vap}}$ is the projection head for the probability of the VAP state. 
In the pretraining, the prediction loss for the VAP state is defined as the cross-entropy loss for the ground-truth class at each time step:
\begin{align}
  \mathcal{L}_{\mathrm{vap}}
  = -\frac{1}{T}\sum_{t=1}^{T}
  \log \hat{p}_{\mathrm{vap},c(t)}(t). \label{eq:vap_loss}
\end{align}
In addition, an auxiliary VAD head $h_{\mathrm{vad}}$ predicts the current speech activity of each speaker.
The VAD head uses each speaker's features as follows:
\begin{align}
  \hat{p}_{\mathrm{vad},u}(t)
  &= h_{\mathrm{vad}}\!\left(\bm{z}_u(t)\right), \\
  \hat{p}_{\mathrm{vad},s}(t)
  &= h_{\mathrm{vad}}\!\left(\bm{z}_s(t)\right).
\end{align}
The VAD loss is defined as follows:
\begin{align}
  \mathcal{L}_{\mathrm{vad}}
  = \frac{1}{2T}\sum_{t=1}^{T}\sum_{a\in\{u,s\}}
  \ell_{\mathrm{bce}}\!\left(y_a(t),\hat{p}_{\mathrm{vad},a}(t)\right).
  \label{eq:vad_loss}
\end{align}
Here, $y_a(t)\in\{0,1\}$ is the ground-truth label for speaker $a$, $\ell_{\mathrm{bce}}$ is the binary cross-entropy loss, and $\hat{p}_{\mathrm{vad},a}(t)$ is the predicted probability.
The original VAP pretraining objective is a weighted sum of these two losses:
\begin{align}
  \mathcal{L}_{\mathrm{tt}}
  = \mathcal{L}_{\mathrm{vap}}
  + w_{\mathrm{vad}}\mathcal{L}_{\mathrm{vad}}.
  \label{eq:turn_taking_loss}
\end{align}

VAP is designed to operate with low computational cost and high processing speed.
However, detecting an event quickly does not necessarily mean that the system can actually output a response at the detected time.
The proposed real-time short-response selection addresses this gap by selecting a response that can be output immediately in parallel with the generation modules.

\section{Task and Method}
\subsection{Latency Mitigation with Real-Time Response Selection}
We introduce real-time short-response selection.
We first define the set of short-response candidates $\mathcal{R}_{s}$ that can be output by the system speaker $s$ as follows:
\begin{align}
  \mathcal{R}_{s}
  = \{\bm{r}_{s,1},\bm{r}_{s,2},\ldots,\bm{r}_{s,N_s}\}
\end{align}
Here, $\bm{r}_{s,n}$ denotes the speech waveform of the $n$-th candidate, and $N_s$ denotes the number of candidates.
These utterances are generated by TTS before the dialogue begins.

Next, we describe the specific processing flow of the proposed model.
We add two heads sharing the streaming feature $\bm{z}(t)$ from Eq.~(\ref{eq:vap_features}):
\begin{align}
  \hat{p}_{\mathrm{tim}}(t)
  &= \sigma\!\left(h_{\mathrm{tim}}(\bm{z}(t))\right),
  \label{eq:timing_head} \\
  \hat{p}_{\mathrm{cls}}(t)
  &= \sigma\!\left(h_{\mathrm{cls}}(\bm{z}(t))\right),
  \label{eq:cls_head}
\end{align}
where $\sigma$ denotes the sigmoid function.
The timing head $h_{\mathrm{tim}}$ detects a turn-taking event at time $t$ when the probability rises above the threshold within one time frame:
\begin{align}
  e(t) =
  \begin{cases}
    1 & \text{if } \hat{p}_{\mathrm{tim}}(t) \geq \theta_{\mathrm{tim}} \text{ and } \hat{p}_{\mathrm{tim}}(t-1) < \theta_{\mathrm{tim}}, \\
    0 & \text{otherwise.}
  \end{cases}
  \label{eq:response_trigger}
\end{align}
Here, $\theta_{\mathrm{tim}}$ is the detection threshold, and $e(t)=1$ indicates a rising edge (i.e., a response should begin one time frame after $t$). The event time $t_e$ is the frame at which $e(t_e)=1$.
In parallel, $\bm{z}(t)$ is fed into the context encoder $g_{\mathrm{ctx}}$ to obtain the dialogue context feature:
\begin{align}
  \bm{h}_{\mathrm{ctx}}(t)
  = g_{\mathrm{ctx}}\!\left(\bm{z}(t)\right)
  \in \mathbb{R}^{D} \label{eq:extract_dialogue_feature}
\end{align}

Each candidate short response is converted before the dialogue begins using the short-response encoder $g_{\mathrm{resp}}$:
\begin{align}
  \bm{h}_{\mathrm{resp},n}
  = g_{\mathrm{resp}}\!\left(\bm{r}_{s,n}\right)
  \in \mathbb{R}^{D},
  \quad n\in\{1,\ldots,N_s\}
\end{align}
When a turn-taking event is detected at time $t_e$, the cosine similarity between the dialogue context feature and each candidate response is computed as:
\begin{align}
  s_n(t_e)
  = \operatorname{cos}\!\left(\bm{h}_{\mathrm{ctx}}(t_e),\bm{h}_{\mathrm{resp},n}\right)
  \label{eq:response_score}
\end{align}
All candidates are then ranked according to this score, and the top-1 candidate is selected:
\begin{align}
  n^{*}(t_e)=\mathop{\arg\max}_{n\in\{1,\ldots,N_s\}}s_n(t_e)
  \label{eq:response_selection}
\end{align}
The selected speech $\bm{r}_{s,n^{*}(t_e)}$ is output at $t_e + 1$ independently of the main generation modules.
Finally, the classification head outputs $\hat{p}_{\mathrm{cls}}(t_c)$ in Eq.~(\ref{eq:cls_head}) at the end of the selected short response $t_c$ to determine whether the main response follows the short response.
Here, $\theta_{\mathrm{cls}}$ is the classification threshold: the main response is judged to follow if $\hat{p}_{\mathrm{cls}}(t_c)\geq\theta_{\mathrm{cls}}$, and otherwise the short response is treated as a backchannel.

\subsection{Proposed Method}
\label{sec:training_objective}
The proposed model is trained with the short-response selection objective in addition to the response timing prediction objective.
For training, we use the start times of main responses and backchannels in two-speaker dialogues, together with their actual response speech.
For main responses, we extract short responses from the first utterance interval after each speaker change by using an external VAD model~\cite{silero_vad}.
For backchannels, we use each utterance interval.

Let $t_i$ denote the start time frame of the $i$-th response.
To select a short response before its onset, we use $\tilde{t}_i=t_i-1$ as the target time, consistent with Eq.~(\ref{eq:response_trigger}).
We pair the dialogue context features at this frame with the positive short response $\bm{r}_i^+$.
For each positive example, we randomly sample $K$ negative examples $\{\bm{r}_{i,k}^-\}_{k=1}^{K}$.
These are sampled from the response pool for the same speaker in the same session.

We use the InfoNCE loss~\cite{infonce} to train short-response selection.
We define the scores for positive and negative examples as follows:
\begin{align}
  s_i^+
  &= \operatorname{cos}\!\left(
    \bm{h}_{\mathrm{ctx}}(\tilde{t}_i),g_{\mathrm{resp}}(\bm{r}_i^+)
  \right), \\
  s_{i,k}^-
  &= \operatorname{cos}\!\left(
    \bm{h}_{\mathrm{ctx}}(\tilde{t}_i),g_{\mathrm{resp}}(\bm{r}_{i,k}^-)
  \right).
\end{align}
The response selection loss is then given as follows:
\begin{align}
  \mathcal{L}_{\mathrm{sel}}
  = -\frac{1}{M}\sum_{i=1}^{M}
  \log\frac{
    \exp(s_i^+/\tau)
  }{
    \exp(s_i^+/\tau)+\sum_{k=1}^{K}\exp(s_{i,k}^-/\tau)
  }.
  \label{eq:selection_loss}
\end{align}
Here, $M$ denotes the number of positive examples in a minibatch and $\tau$ the temperature parameter.

Response timing prediction is realized by combining the following two subtasks.
The first subtask predicts when a response should begin.
Let $y_{\mathrm{tim}}(t)\in[0,1]$ denote the training target.
We use the binary cross-entropy loss $\ell_{\mathrm{bce}}$ for timing prediction.
The loss is given as follows:
\begin{align}
  \mathcal{L}_{\mathrm{tim}}
  &= \frac{1}{T}\sum_{t=1}^{T} \ell_{\mathrm{bce}}\!\bigl(
    y_{\mathrm{tim}}(t), \hat{p}_{\mathrm{tim}}(t)
  \bigr).
  \label{eq:timing_loss}
\end{align}

The second subtask is event classification conditioned on a rising edge in the timing prediction.
The training target $y_{\mathrm{cls}}(t)$ is 1 for main responses and 0 for backchannels.
The interval used for training is $[t_i,e_i)$.
The interval end is $e_i=t_i+\min(f_{\mathrm{r}},L_i)$, where $f_{\mathrm{r}}$ is the frame rate (frames per second), so that $f_{\mathrm{r}}$ frames correspond to one second, and $L_i$ is the short-response length measured in the time frames.
The event classification loss is given as follows:
\begin{align}
  \mathcal{L}_{\mathrm{cls}}
  &= \frac{1}{\lvert\mathcal{T}_{\mathrm{cls}}\rvert}\sum_{t\in\mathcal{T}_{\mathrm{cls}}} \ell_{\mathrm{bce}}\!\bigl(
    y_{\mathrm{cls}}(t), \hat{p}_{\mathrm{cls}}(t)
  \bigr).
  \label{eq:is_main_loss}
\end{align}
Here, $\mathcal{T}_{\mathrm{cls}}$ denotes the set of target frames.

The total loss is a weighted sum of the short-response selection, timing prediction, and event classification losses:
\begin{align}
  \mathcal{L}_{\mathrm{total}}
  ={}& \lambda_{\mathrm{sel}}\mathcal{L}_{\mathrm{sel}}
  + \lambda_{\mathrm{tim}}\mathcal{L}_{\mathrm{tim}} + \lambda_{\mathrm{cls}}\mathcal{L}_{\mathrm{cls}}.
  \label{eq:total_training_loss}
\end{align}
Here, $\lambda_{\mathrm{sel}}$, $\lambda_{\mathrm{tim}}$, and $\lambda_{\mathrm{cls}}$ are the loss weights.

\section{Experimental Evaluation}
\subsection{Experimental Setup}
\newcommand{\scoreblank}{\underline{\hspace{1.5em}}}
\begin{table}[t]
  \caption{Total duration and numbers of sessions, unique speakers, and target responses in each partition.}
  \label{tab:data_partitions}
  \centering
  \setlength{\tabcolsep}{2pt}
  \begin{tabular}{lrrrr}
    \hline
    Partition & Sessions & Speakers & \shortstack{Target\\responses} & Duration [h] \\
    \hline
    Train & 195 & 170 & 98,329 & 69.09 \\
    Valid & 14 & 10 & 8,094 & 4.52 \\
    Test & 38 & 31 & 18,103 & 12.19 \\
    \hline
  \end{tabular}
\end{table}

For training and evaluation, we extracted two-speaker dialogues from the Corpus of Everyday Japanese Conversation (CEJC)~\cite{cejc-corpus}.
For each dialogue, we used separate audio recordings and utterance interval annotations.
We grouped sessions by recording contributor ID and divided the groups into training, validation, and test sets with target proportions of $80$\%, $5$\%, and $15$\%, respectively, so that each group was assigned to only one set.
Table~\ref{tab:data_partitions} shows the total duration and the numbers of sessions, unique speakers, and utterances targeted for short-response extraction in each set.
The utterance counts include both speakers.
Since either speaker in a dialogue can play the system role, we randomly swapped the user and system roles of the two speakers with a probability of $50$\% during training, and evaluated both role assignments for each dialogue at test time.

We created response labels by processing utterance intervals in order of their start times.
For an utterance by a speaker other than the current main speaker, we labeled it as a backchannel if it ended within $1.0$ second of the end of the most recent main utterance, and otherwise as a main response, which was treated as a speaker change.
For short-response selection, we used these backchannels and the first main response after each speaker change as positive examples, detecting the first speech segment in each target utterance with Silero VAD~\cite{silero_vad} and extracting a continuous segment of up to $2.0$ seconds as the short response.

We initialized the turn-taking model with a pretrained Japanese VAP model~\cite{noise-robust-vap,noise-robust-vap-hf}, using the $10$ Hz version with a context length of $3.0$ seconds for each attention module.
For the response encoder, we concatenated features pre-extracted from layers $6$, $18$, and $24$ of the Japanese HuBERT model~\cite{hubert,hubert_aist_large} at each time step to obtain $3\text{,}072$-dimensional features, projected them to $512$ dimensions, and fed them into a three-layer Transformer.
The outputs were averaged over valid frames and projected to $128$ dimensions.
The context encoder $g_{\mathrm{ctx}}$ consisted of a three-layer MLP with a hidden dimension of $512$ and a projection layer to $128$ dimensions.
Dropout~\cite{dropout} with a probability of $0.1$ was applied to all heads in the model.

We froze the CPC~\cite{cpc_encoder} extractor and HuBERT, the front ends of the turn-taking model and response encoder, respectively.
We trained the VAP model, the two binary heads, and the short-response encoders.
In preliminary experiments, propagating the response selection loss to the turn-taking model slightly degraded the timing prediction performance.
We therefore detached the VAP features in Eq.~(\ref{eq:extract_dialogue_feature}) used for response selection, so that the response selection loss does not update the turn-taking model.
For response selection, we used four negative examples per positive example and set the InfoNCE temperature to $\tau=0.07$.

In preliminary experiments with actual model inference, dialogue breakdowns frequently occurred because the prediction probabilities failed to rise during long silences.
To prevent this, we applied silence augmentation to the timing prediction task during training only.
For each main response and backchannel of the system speaker that starts within a training segment, we replaced its speech with silence with a probability of $50$\%, independently for each event.
For the silenced events, the timing target was set to 1 from the response onset until the end of the original utterance or $2.0$ seconds after the onset, whichever came first, so that the model keeps predicting the response while no speech is observed.
The event classification loss was excluded over the silenced utterance interval.

To address data imbalance on the timing prediction task, we applied two techniques.
First, we assigned soft labels of $0.2, 0.4, 0.6$, and $0.8$ to the four frames immediately preceding the ground-truth label.
Second, we applied inverse-frequency weighting~\cite{imbalanced_classification} to the loss in Eq.~(\ref{eq:timing_loss}).
The positive-class weight was approximately 6.50.
The loss weights in Eq.~(\ref{eq:total_training_loss}) were $\lambda_{\mathrm{sel}}=5.0$, $\lambda_{\mathrm{tim}}=1.0$, and $\lambda_{\mathrm{cls}}=1.0$.
We used AdamW with a maximum learning rate of $5\times10^{-5}$, increasing it linearly from $1\times10^{-10}$ during the first $10$\% of all update steps and then keeping it constant.
We used a batch size of $16$, updated the parameters after each batch, and trained for $10$ epochs using mixed precision with `bfloat16'.

\subsection{Results}
\subsubsection{Response Selection Performance}
\begin{table}[t]
  \caption{
    Text Top-1 accuracy (\%) of short-response selection on 14,446 test events.
    A selection is correct when its text matches that of the target response.
    BC denotes backchannel.
  }
  \label{tab:response_selection_results}
  \centering
  \setlength{\tabcolsep}{6pt}
  \begin{tabular}{lrrr}
    \hline
    Type & Chance rate & Proposed & LLM \\
    \hline
    Main & 0.54 & 4.52 & 1.62 \\
    BC & 8.39 & 19.51 & 9.55 \\
    All & 6.41 & 15.72 & 7.55 \\
    \hline
  \end{tabular}
\end{table}
\begin{table}[t]
  \caption{
    Speech ranking accuracy (\%) of the proposed method on 14,446 test events.
    Top-$k$\% is the rate at which the target speech ranks within the top $k$\% of the candidate pool.
    The chance rate of Top-$k$\% is $k$\%. BC denotes backchannel.
  }
  \label{tab:speech_ranking_results}
  \centering
  \setlength{\tabcolsep}{4pt}
  \begin{tabular}{lrrrrr}
    \hline
    Type & 1\% & 5\% & 10\% & 25\% & 50\% \\
    \hline
    Main & 7.55 & 24.32 & 39.25 & 62.17 & 83.98 \\
    BC & 4.72 & 18.50 & 31.55 & 58.60 & 83.23 \\
    All & 5.41 & 19.97 & 33.50 & 59.50 & 83.42 \\
    \hline
  \end{tabular}
\end{table}
\begin{table}[t]
  \caption{
    Number of response candidates per test event ($N=14{,}446$).
  }
  \label{tab:response_candidate_counts}
  \centering
  \setlength{\tabcolsep}{3pt}
  \begin{tabular}{lrrr}
    \hline
    Candidate type & Min & Max & $\mu \pm \sigma$ \\
    \hline
    Speech & 29 & 985 & $365.71 \pm 262.09$ \\
    Text & 23 & 488 & $182.63 \pm 106.90$ \\
    \hline
  \end{tabular}
\end{table}
We evaluated response selection performance using two metrics.
Top-1 accuracy (text) counts a selection as correct when the text of the selected candidate matches that of the target response, because candidates with the same text cannot be distinguished by the LLM, which takes only text as input.
Top-$k$\% accuracy~\cite{contrastive_bc_selection} measures whether the target speech itself ranks within the top $k$\% of the speech candidate pool.
Since the LLM outputs a single candidate ID and no ranking, we report Top-$k$\% only for the proposed method, and Top-1 accuracy (text) is the only metric comparable between the two methods.
We adopted two baseline methods.
One is the chance rate, which represents the expected value when responses are selected at random.
The other is LLM-based selection, inspired by Liu et al.~\cite{llm_based_selection}.
We used Gemma 4 E2B-it~\cite{gemma4} as the LLM for response selection.
The English prompt~\footnote{The actual prompts and examples were given in Japanese.} for selecting the short response is described as follows:
\begin{quote}
\footnotesize
\texttt{[Example 1, user message]}\\
Recent conversation:\\
Other speaker: \texttt{[utterance 1]}\\
Other speaker: \texttt{[utterance 2]}\\
Responder: \texttt{[utterance 3]}\\
\texttt{[...]}\\
Candidate IDs and utterances:\\
\texttt{128: [candidate transcription]}\\
\texttt{135: [candidate transcription]}\\
\texttt{[...]}\\
\texttt{[same selection instruction as below]}\\
\texttt{[Example 1, assistant message]}\\
\texttt{135}\\[0.5em]
\texttt{[Example 2, user: same format]}\\
\texttt{[Example 2, assistant: correct ID]}\\[0.5em]
\texttt{[Test, user message]}\\
Recent conversation:\\
\texttt{[history]}\\[0.5em]
Candidate IDs and utterances:\\
\texttt{[options]}\\[0.5em]
Considering the conversation above, select the ID of the most appropriate candidate as the \texttt{[response]} from the given response candidates. Reply with only the numeric candidate ID.
\end{quote}
Here, \texttt{[history]} contains up to eight recent utterances labeled as the other speaker or responder.
\texttt{[options]} lists candidate IDs and their text sequences, one per line.
For backchannels, we used the ground-truth sequences provided in the dataset, 
and for main responses, we used the ASR transcriptions of the short responses. 
We also removed duplicate short responses in the text domain.
Table~\ref{tab:response_candidate_counts} summarizes the statistics of the short-response candidate pools. 
For \texttt{[response]}, we used ``next response'' for a main response or ``backchannel to the current utterance'' for a backchannel.

Tables~\ref{tab:response_selection_results} and \ref{tab:speech_ranking_results} show the text Top-1 accuracy and the speech ranking results, respectively.
Comparing the proposed method with the chance rate shows that the proposed method achieved higher scores on all metrics. 
This result suggests that the proposed method selects responses based on the dialogue context. 
The proposed method also achieved higher scores than LLM-based selection. 
Our analysis showed that the LLM tended to favor candidates near the beginning of the list. 
When the number of candidates was large, this selection bias~\cite{selection_bias1,selection_bias2} may have adversely affected the LLM's performance.
The proposed method is not affected by the order in which candidates are presented, making it a better choice in this respect.

\subsubsection{Impact on Turn-Taking Prediction}
\begin{table}[t]
  \caption{
    Event-wise response-onset timing scores (\%) and mean absolute error (MAE). 
    $\epsilon$ is the absolute timing tolerance in seconds.
  }
  \label{tab:timing_results}
  \centering
  \setlength{\tabcolsep}{1.5pt}
  \begin{tabular}{llrrr|r}
    \hline
    & & \multicolumn{3}{c|}{$\epsilon$ (s)} & \\
    Metric & Method & 0.2 & 0.5 & 1.0 & MAE (s) \\
    \hline
    Precision & Timing only        & 60.81 & 78.14 & 88.68 & - \\
              & Proposed           & 67.04 & 82.45 & 91.15 & - \\
    \hline
    Recall & Timing only        & 54.83 & 78.48 & 90.93 & - \\
           & Proposed           & 50.35 & 74.42 & 88.31 & - \\
    \hline
    F1 & Timing only        & 57.66 & 78.31 & 89.79 & 0.488 \\
       & Proposed           & 57.51 & 78.23 & 89.71 & 0.477 \\
    \hline
  \end{tabular}
\end{table}

\begin{table}[t]
  \caption{Event-classification F1 scores (\%). BC is the event of backchannels. }
  \label{tab:response_type_results}
  \centering
  \setlength{\tabcolsep}{5pt}
  \begin{tabular}{lrrr}
    \hline
    Method & Main & BC & Macro \\
    \hline
    Timing only & 79.95 & 85.73 & 82.84 \\
    Proposed  & 79.61 & 86.20 & 82.90 \\
    \hline
  \end{tabular}
\end{table}
We conducted an objective evaluation of the timing prediction to confirm that adding response selection does not degrade its performance.
To assess timing prediction accuracy, we used the following metrics focused on prediction triggers.
Let $\epsilon$ denote the timing tolerance.
We count the first trigger as $N_{\mathrm{on}}$ if it falls within $\pm\epsilon$ of the target.
Earlier, later, and missing triggers are counted as $N_{\mathrm{early}}$, $N_{\mathrm{late}}$, and $N_{\mathrm{miss}}$, respectively.
We define precision, recall, and F1 as follows:
\begin{align}
  P &= \frac{N_{\mathrm{on}}}{N_{\mathrm{on}}+N_{\mathrm{early}}}, \\
  R &= \frac{N_{\mathrm{on}}}{N_{\mathrm{on}}+N_{\mathrm{late}}+N_{\mathrm{miss}}}, \\
  F_1 &= \frac{2PR}{P+R}.
\end{align}
Additionally, MAE is computed relative to the shifted target time over events with a detected trigger.

Table~\ref{tab:timing_results} shows the timing prediction results.
The proposed method achieved F1 scores comparable to those of the model trained without the response-selection objective, 
while showing a slightly better MAE.
These results indicate that timing prediction performance can be maintained while incorporating response selection.
The substantial differences in precision and recall between the two methods arise from the different operating points selected by maximizing the F1 score on the validation set.

Next, for event classification, we used the ground-truth onset times in the test set as the reference.
We evaluated all $113\text{,}247$ frames from response onset until the response ended or $1.0$ second had elapsed.
Table~\ref{tab:response_type_results} shows the F1 scores for main responses and backchannels, along with macro F1.
These results show that integrating response selection did not worsen overall event classification performance.

\subsubsection{Processing Time}
We evaluated the speed of the proposed response selection method to assess its real-time performance.
The evaluation covered the response selection process, the response prediction module in the preface-generation baseline~\cite{context_aware_preface_generation}, and the VAP model~\cite{noise-robust-vap-hf}, which serves as the backbone of the proposed method.
For reference, we also measured streaming ASR~\cite{streaming_nemotron} under the same conditions.
Each measurement was repeated 30 times on a MacBook Air equipped with an Apple M5 chip.

\begin{table}[t]
  \caption{
    Processing times of each method.
    The upper and lower sections show measurements on the CPU and Metal Performance Shaders (MPS), respectively.
    $N$ denotes the number of selection candidates.
    The proposed method operates at 10 Hz, and Nemotron ASR operates at 3.125 Hz.
    The preface-generation result covers only the response readiness estimation module.
    Note that subsequent LLM and TTS processing adds further latency in practice.
  }
  \label{tab:processing_times}
  \centering
  \setlength{\tabcolsep}{1.25pt}
  \begin{tabular}{lrr}
    \hline
    Case & $\mu \pm \sigma$ (ms) & P95 (ms) \\
    \hline
    Proposed & $9.3749 \pm 0.0672$ & 9.4883 \\
    \hspace{0.5em}+Select. ($N=10^{1}$) & $0.0231 \pm 0.0007$ & 0.0246 \\
    \hspace{0.5em}+Select. ($N=10^{2}$) & $0.0281 \pm 0.0008$ & 0.0292 \\
    \hspace{0.5em}+Select. ($N=10^{3}$) & $0.0332 \pm 0.0009$ & 0.0347 \\
    \hspace{0.5em}+Select. ($N=10^{4}$) & $0.0765 \pm 0.0009$ & 0.0785 \\
    Preface-gen.~\cite{context_aware_preface_generation} & $4.3265 \pm 0.1175$ & 4.6139 \\
    \hline
    \noalign{\vskip 3pt}
    Nemotron ASR~\cite{streaming_nemotron} & $36.6975 \pm 5.9689$ & 46.8798 \\
    \multicolumn{3}{l}{LLM select. (Gemma 4 E2B)~\cite{gemma4}} \\
    \hspace{0.5em}$N=10^{1}$ & $249.4189 \pm 1.1605$ & 250.5442 \\
    \hspace{0.5em}$N=10^{2}$ & $718.4107 \pm 0.6725$ & 719.4658 \\
    \hspace{0.5em}$N=10^{3}$ & $(5.66 \pm 0.37) \times 10^{4}$ & $6.33 \times 10^{4}$ \\
    \hspace{0.5em}$N=10^{4}$ & \multicolumn{2}{c}{Out-of-memory} \\
    \hline
  \end{tabular}
\end{table}

Table~\ref{tab:processing_times} shows the speed evaluation results.
The VAP model runs in 9 ms per inference, well within the 100 ms available at 10 Hz.
The response selection process adds almost no overhead, even with $10\text{,}000$ candidates.
These results show that our method can be introduced without compromising the real-time performance of the turn-taking model.
Unlike LLM-based filler selection~\cite{llm_based_selection}, our method requires no online LLM inference for selecting a short response.
The selection step added only $0.076$ ms with $10\text{,}000$ candidates on the evaluated CPU.
Preface-generation also runs very quickly, but it relies on recognized text and is therefore affected by the ASR process.
Moreover, because it requires subsequent generation by an LLM and TTS system, the size of the modules that can be used is limited.
Therefore, our method can select a response with minimal overhead, independently of ASR, LLM, and TTS latency.

\section{Discussion}
\noindent\textbf{One-to-many mapping between dialogue contexts and short responses.}
Our method learns to associate dialogue context features based on causal inputs to the VAP model with short-response features.
As shown in Table~\ref{tab:response_selection_results}, the results demonstrate the effectiveness of this approach.
However, the model has a limitation in capturing real dialogue.
A causal input does not always determine a unique response.
For example, several responses may be appropriate to ``What do you think?''
The choice may remain open until the speaker's intent becomes clear.
Our method does not account for this one-to-many mapping.
Addressing this limitation is an important direction for future work.

\noindent\textbf{Consistency between the two output paths.}
A major limitation of our method is that the selected short response can be inconsistent with the main response generated afterward.
For example, dialogue quality can suffer if our method selects a positively toned response and the LLM then generates a negative one.
One possible approach is to provide the context of the selected response to the main LLM.
However, persona information is generally handled by the main LLM, so having a smaller LLM used for response selection~\cite{llm_based_selection} determine the response strategy is not ideal.
We leave the development of a method that improves consistency with the main response while preserving real-time performance to future work.

\noindent\textbf{Domain mismatch between training and inference.}
Our method is trained on short responses extracted from real conversations, whereas synthetic speech is used for short responses at inference time.
Differences in acoustic features, such as prosody, and in the distribution of the short-response pool can lead to degraded performance.
Future work should explore approaches such as post-training with a reward model~\cite{rlaif} to mitigate this mismatch.

\section{Conclusion}
We proposed a real-time short-response selection task for cascaded spoken dialogue systems.
The method uses contrastive learning to associate streaming VAP features with short spoken responses.
It selects pre-generated speech independently of ASR, LLM, and TTS processing.
Experiments on CEJC showed better selection performance than random ranking.
Timing prediction and overall event classification performance remained comparable to those without response selection.
On the evaluated Apple M5 CPU, selection took only $0.076$ ms with $10,000$ candidates.
These results support integrating short-response selection into turn-taking models with minimal processing overhead.

\section*{Acknowledgments}
A part of this research was conducted by AIST and sub-contract entity, 
The University of Osaka, under FRONTia, a Japanese national program led by the Ministry of Economy, 
Trade and Industry of Japan (METI) and the New Energy and Industrial Technology Development Organization (NEDO), 
aimed at developing a domestic multimodal foundation model for AI robots and physical AI.

\section*{Declaration of generative AI use}
Generative AI tools (Claude Code and Codex) were used as supplementary aids for English proofreading and for implementing the experimental code.
The authors reviewed all outputs and take full responsibility for the content of this paper.

\bibliographystyle{ast}
\bibliography{ref}

\end{document}